\documentclass[fleqn,twoside]{article}
\usepackage{espcrc2}

\usepackage{graphicx}

\title{Finite volume effects using lattice chiral perturbation theory}

\author{Bu\=gra Borasoy\address{Physik Department, Technische Universit\"at
                                M\"unchen, D-85747 Garching, Germany},
        Randy Lewis\address[Regina]{Department of Physics, University of
                                    Regina, Regina, SK, S4S 0A2, Canada}
        and
        Daniel Mazur\addressmark[Regina]}
       
\begin{document}

\begin{abstract}
Lattice regularization is used to perform chiral perturbation theory
calculations in finite volume.  The lattice spacing is chosen small enough
to be irrelevant, and numerical results are obtained from simple summations.
\end{abstract}

\maketitle

\section{CONTEXT}

Lattice QCD simulations are necessarily performed in a finite volume
though one is typically interested in results for infinite spacetime.
The extrapolation introduces a systematic uncertainty which needs to be
controlled.
As the lightest particle in the QCD spectrum, the pion has a large Compton
wavelength and therefore plays a key role in these volume
effects, so chiral perturbation theory is the natural
tool for studying volume dependences.  After the pioneering work of Gasser
and Leutwyler\cite{GL88}, there has been a lot of activity on this topic.  Some
recent studies in the light meson sector can be found in
Refs.~\cite{recent,ColDur,ColHae,Col}.

Physical results do not depend on regularization scheme.
For chiral perturbation theory in a finite volume,
lattice regularization is numerically convenient because
loop diagrams are finite summations for any nonzero lattice spacing.
Divergences would appear as the lattice spacing vanishes, but
for the volume effects studied here
we can simply use a nonzero lattice spacing which is small enough
to be numerically irrelevant.

The present work uses the lattice regularized chiral perturbation theory
Lagrangian of Ref.~\cite{lchpt}, but for two flavors rather than three.
Extra $O(a)$ terms could be added to the
Lagrangian but they are irrelevant in the continuum limit, and our present goal
is the computation of finite volume effects for the continuum limit.

\section{THE PION MASS}

\begin{figure*}[htb]
\includegraphics[width=14cm]{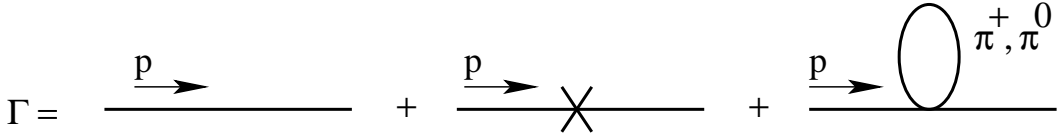}
\vspace{-8mm}
\caption{One-loop Feynman diagrams for the pion mass.}
\label{massfeyn}
\end{figure*}

Consider an isotropic spacetime lattice with spacing ``$a$'',
$N_s$ sites in each spatial direction and $N_t$ sites in the temporal
direction.  The pion mass is obtained at the one-loop level from the diagrams
in Fig.~\ref{massfeyn}.  The propagators and vertices are obtained from the
Lagrangian of Ref.~\cite{lchpt} in a straightforward manner, and loop momenta
are summed.  The resulting Green's function is
\begin{eqnarray}
\Gamma &=& \Gamma_{\rm LO} + \Gamma_{\rm NLO}^{(a)} + \Gamma_{\rm NLO}^{(b)},
    \label{Gammaeqn} \\
\Gamma_{\rm LO} &=& -x_\pi^2 - \frac{2}{a^2}\sum_\mu(1-\cos ap_\mu), \\
\Gamma_{\rm NLO}^{(a)} &=& -\frac{2}{3a^4F^2} - \frac{2x_\pi^4}{F^2}(l_3+l_4)
           \nonumber \\
           &&              - \frac{2l_4x_\pi^2}{a^2F^2}\sum_\mu\sin^2ap_\mu, \\
\Gamma_{\rm NLO}^{(b)} &=& \frac{1}{6N_s^3N_ta^4F^2}\sum_k
           \bigg(112+5a^2x_\pi^2 \nonumber \\
           && -20\sum_\mu\cos ap_\mu-20\sum_\mu\cos ak_\mu \nonumber \\
           && \left. +12\sum_\mu\cos ap_\mu\cos ak_\mu\right)D(k),
\end{eqnarray}
where
\begin{equation}
D(k) = \frac{1}{a^2x_\pi^2+2\sum_\mu(1-\cos ak_\mu)}
\end{equation}
is the pion propagator and
\begin{equation}
x_\pi = \sqrt{2Bm_q}
\end{equation}
is the lowest-order pion mass in the continuum limit.  (We work in the isospin
limit, $m_q\equiv m_u=m_d$.)
The momentum summations include $k_{1,2,3}=\frac{2\pi n}{aN_s}$
with $n=1, 2, 3, \ldots N_s$
and $k_4=\frac{2\pi n}{aN_t}$ with $n=1, 2, 3, \ldots N_t$.

The pion mass is obtained by solving $\Gamma=0$ for $ip_4$ with $\vec p=0$.
The result is
\begin{eqnarray}
M_\pi &=& \frac{2}{a}{\rm arcsinh}\left(\frac{aX_\pi}{2}\right), \\
X_\pi^2 &=& x_\pi^2 + \frac{2x_\pi^4l_3}{F^2} \nonumber \\
        && +x_\pi^2\sum_k\frac{(3-2\cos ak_4)}{2N_s^3N_ta^2F^2}D(k) + O(a).
\label{masseqn}
\end{eqnarray}
Having simply written down the pion mass in terms of Feynman vertices and
propagators, we can now do the summation numerically to obtain the pion mass.
The numerical values of the parameters $x_\pi$ and $l_3$ will depend
sensitively on lattice spacing since they must absorb terms that diverge as
$a\to0$.  Nevertheless the computation is finite for any $a\neq0$, and
for sufficiently small $a$ the observable pion mass is independent of $a$.

\begin{figure}[htb]
\includegraphics[width=7cm]{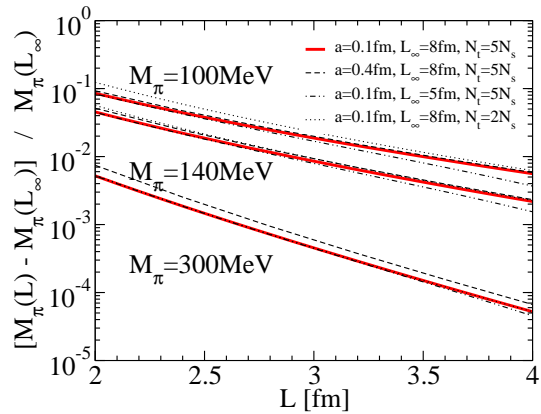}
\vspace{-8mm}
\caption{Volume effects for the pion mass.}
\label{massplot}
\vspace{-3mm}
\end{figure}

To study spatial volume effects on a lattice of infinite temporal extent,
we can choose $N_t\gg N_s$ and compute the {\em difference} of $M_\pi$ at
two different spatial volumes.  In this computation, the $l_3$ term and the
leading $x_\pi^2$ term and all ``would-be divergences'' subtract away.
The resulting computation of volume dependence is plotted in
Fig.~\ref{massplot}.  By choosing $a=0.1$ fm, $N_t=5N_s$ and approximating
an infinite volume by $L_s=8$ fm, the numerical results are found to agree
with the standard one-loop results
discussed, for example, in Ref.~\cite{ColDur}.  Figure \ref{massplot} also
shows the
errors that would arise by increasing $a$, decreasing $N_t$ or decreasing
the approximation to infinite spatial length.

It is clear from the simplicity of Eq.~(\ref{masseqn}) that the computation
is inexpensive.

\section{OTHER PION OBSERVABLES}

\begin{figure*}[htb]
\includegraphics[width=135mm]{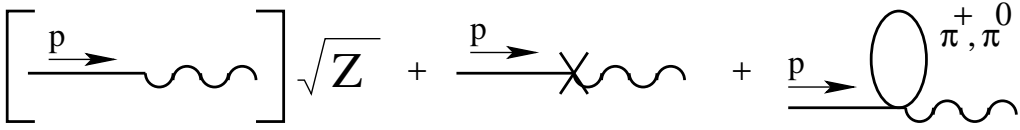}
\vspace{-8mm}
\caption{One-loop Feynman diagrams for the pion decay constant.}
\label{fpifeyn}
\end{figure*}

\begin{figure*}[htb]
\includegraphics[width=135mm]{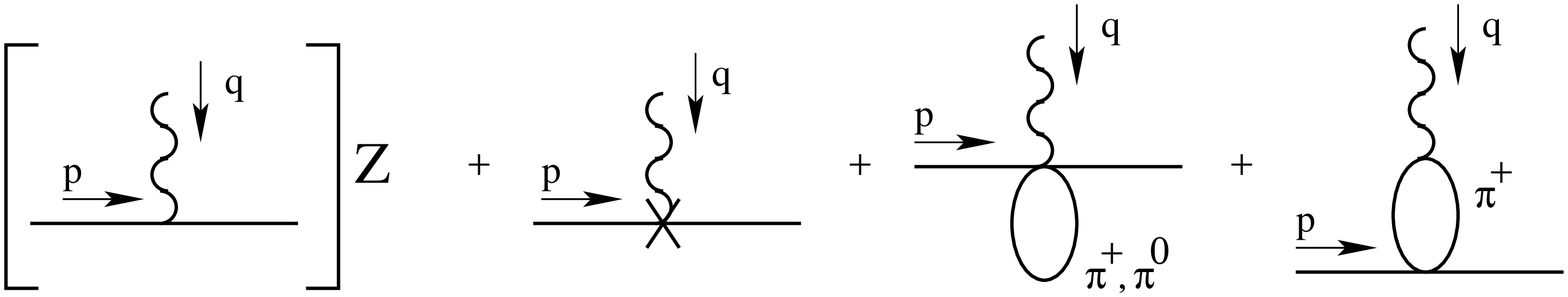}
\vspace{-8mm}
\caption{One-loop Feynman diagrams for the pion form factor.}
\label{pionfffeyn}
\end{figure*}

\begin{figure}[htb]
\vspace{+1mm}
\includegraphics[width=7cm]{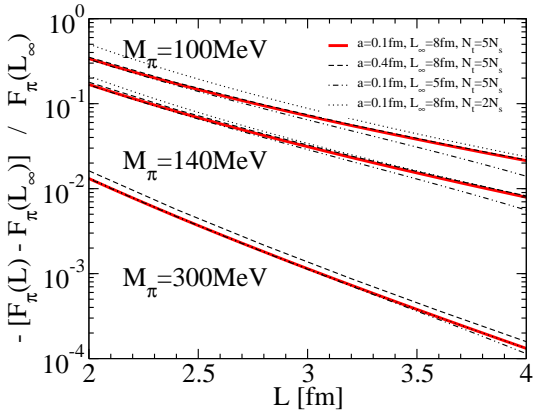}
\vspace{-8mm}
\caption{Volume effects for the pion decay constant.}
\label{fpiplot}
\end{figure}

\begin{figure}[htb]
\vspace{-4mm}
\includegraphics[width=7cm]{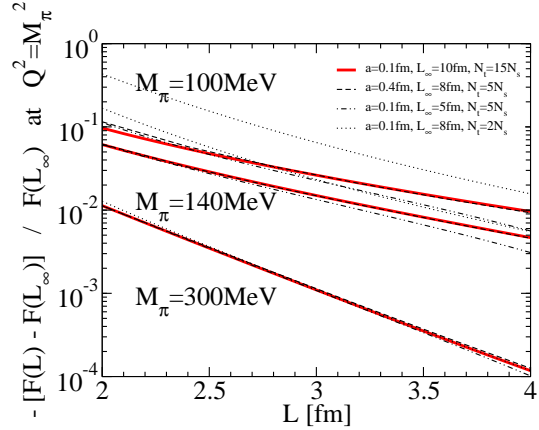}
\vspace{-8mm}
\caption{Volume effects for the pion form factor.}
\label{pionffplot}
\end{figure}

\begin{figure}[htb]
\vspace{1mm}
\includegraphics[width=7cm]{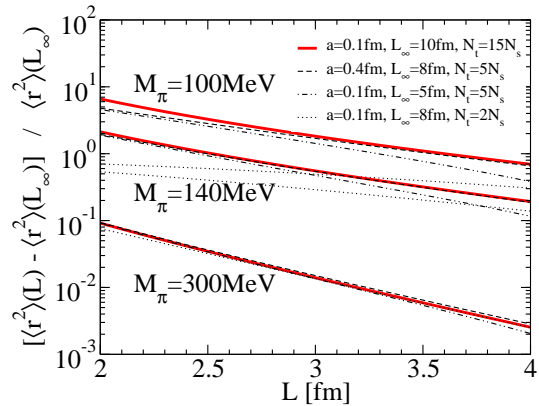}
\vspace{-8mm}
\caption{Volume effects for the pion charge radius.}
\label{rsqplot}
\end{figure}

Other observables can be computed similarly.  The Feynman
diagrams for the pion decay constant and form factor are shown in
Figs.~\ref{fpifeyn} and \ref{pionfffeyn} respectively, where $Z$ denotes
wave function renormalization and is obtained from Eq.~(\ref{Gammaeqn})
in the usual way.  Writing down the vertices and propagators of
Figs.~\ref{fpifeyn} and \ref{pionfffeyn} with a summation over each loop
momentum leads
directly to finite expressions that can be computed numerically.
The volume dependences of the decay constant, form
factor and charge radius (obtained by differentiating the form factor)
are shown in Figs.~\ref{fpiplot}, \ref{pionffplot} and \ref{rsqplot}
respectively.  Results for the decay constant agree with the one-loop
calculation of Ref.~\cite{ColHae}.
Notice that the charge radius has a large fractional dependence
on volume since loop effects occur at
leading order for this observable.

\vspace{3mm}
\noindent
{\bf Acknowledgements}

This work was supported in part by Deutsche Forschungsgemeinschaft,
the Natural Sciences and Engineering Research Council of Canada, and the
Canada Research Chairs Program.

\end{document}